\documentclass[aps,twocolumn,showpacs,floats]{revtex4}
\usepackage{graphicx}
\usepackage{dcolumn}
\usepackage{bm}
\usepackage{color}
\usepackage{amsmath}
\usepackage{amssymb}
\usepackage{hyperref}

\begin{document}

\title{Ferromagnetism in a Repulsive Atomic Fermi Gas with Correlated Disorder}

\author{S. Pilati}
\author{E. Fratini}
\affiliation{The Abdus Salam International Centre for Theoretical Physics, 34151 Trieste, Italy}

\begin{abstract}
We investigate the zero-temperature ferromagnetic behavior of a two-component repulsive Fermi gas in the presence of a correlated random field that represents an optical speckle pattern.
The density is tuned so that the (noninteracting) Fermi energy is close to the mobility edge of the Anderson localization transition. We employ quantum Monte Carlo simulations to determine various ground-state properties, including the equation of state, the magnetic susceptibility, and the energy of an impurity immersed in a polarized Fermi gas (repulsive polaron).
In the weakly interacting limit, the magnetic susceptibility is found to be suppressed by disorder. However, it rapidly increases with the interaction strength, and it diverges at a much weaker interaction strength compared to the clean gas. Both the transition from the paramagnetic phase to the partially ferromagnetic phase, and the one from the partially to the fully ferromagnetic phase are strongly favored by disorder, indicating a case of order induced  by disorder.
\end{abstract}

\maketitle


Understanding the phenomena induced by the interactions in fermionic systems exposed to strong enough disorder to cause the Anderson localization of the single-particle states~\cite{anderson1958absence} (a regime which has been referred to as Fermi glass~\cite{freedman,fleishman}) is one of the most relevant problems in condensed matter physics~\cite{benenti,sanpera,byczuk2005mott}.
Following the first observations of the Anderson localization of matter waves~\cite{billy2008direct,roati2008anderson,jendrzejewski2012three,kondov2011three}, the experiments performed with ultracold atoms exposed to optical speckle patterns have emerged as the ideal platform to explore the intricate interplay between disorder and interactions in a controllable setup~\cite{aspect2009anderson,sanchez2010disordered}. Not only can experimentalists tune the interaction strength~\cite{chin}, but they can also control the disorder amplitude and manipulate its spatial correlations~\cite{mcgehee2013three}.
Some recent theoretical and computational advancements have allowed scientists to precisely determine the mobility edge (namely, the energy threshold separating the localized single-particle orbitals from the extended ones) using realistic models of the speckle pattern~\cite{delande2014mobility,fratini1,fratini2}, thus paving the way to a quantitative comparison with accurate experimental measurements~\cite{semeghini}.
%
%
\begin{figure}
\begin{center}
\includegraphics[width=1.0\columnwidth]{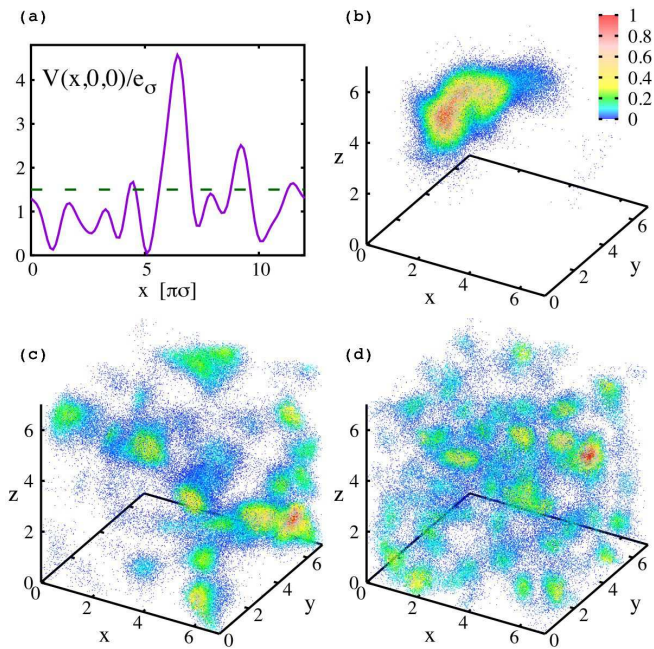}
\caption{(Color online) Panel (a):  speckle pattern intensity $V(x,0,0)/e_\sigma$ along a one-dimensional cross-section. $e_\sigma=\hbar^2/(m\sigma^2)$ is the disorder correlation energy. The length unit is the disorder correlation length $\pi\sigma$. The dashed (green) line indicates the average intensity $V_{\mathrm{dis}}=1.5e_\sigma$.
Panels (b,c,d): particle density distribution for three representative single-particle eigenstates $\phi_j({\bf r})$ of a speckle pattern:  (b) ground-state ($j=0$); (c) state at the Fermi energy $e_j=e_F\simeq 0.84e_\sigma$ corresponding to the density $n=N/L^3\cong0.185(\pi\sigma)^{-3}$ (notice that the mobility edge is $e_{\mathrm{c}}=0.80(3)e_{\sigma}$); (d) state at the energy $e_j\simeq1.2e_\sigma$. The particle density is proportional to the probability density $\left|\phi_j({\bf r})\right|^2$, with the normalization set so that $\mathrm{max}\left(\left|\phi_j({\bf r})\right|^2\right)=1$; the color scale indicates the probability density.}
\label{fig1}
\end{center}
\end{figure}
%
%
While these previous theoretical studies have addressed systems of noninteracting particles, in this Rapid Communication we employ quantum Monte Carlo simulations to investigate the zero-temperature properties of disordered \emph{and} interacting Fermi gases. In particular, we consider a two-component mixture with short-range repulsive interspecies interactions, which is exposed to a blue-detuned isotropic optical speckle pattern~\cite{goodman1975statistical,goodman2007speckle}. We model this system using a realistic continuous-space Hamiltonian that takes into account the spatial correlations of the speckles.\\
Our main interest is to inspect what impact the disorder has on the so-called Stoner instability~\cite{stoner}, namely, the ferromagnetic transition which is supposed to occur in clean Fermi gases when the interatomic repulsion becomes sufficiently strong.
The Stoner instability is one of the standard paradigms in the theory of quantum magnetism. It was proposed as the minimal model to explain itinerant ferromagnetism in certain transition metals. Being a strong-interaction phenomenon, its nature and even its subsistence are still controversial. So far, in solid state systems it has not been possible to unambiguously identify the Stoner mechanism due to the presence of complicated band structures, of disorder, and due to the lack of control over the interaction strength.
Instead, the results of a very recent cold-atoms experiment~\cite{roati} (in which the problems related to the three-body recombinations~\cite{jo2009itinerant,ketterle2012,ketterle2012prl} have been circumvented by preparing a configuration with fully separated components~\cite{zintchenko}) are consistent with the spin fluctuations expected in the vicinity of the Stoner instability~\cite{recati2011spin} and with the quantum Monte Carlo predictions for the critical repulsion strength in clean systems~\cite{conduit,pilati2010,trivedi}.\\
%
%
%
\begin{figure}
\begin{center}
\includegraphics[width=1.0\columnwidth]{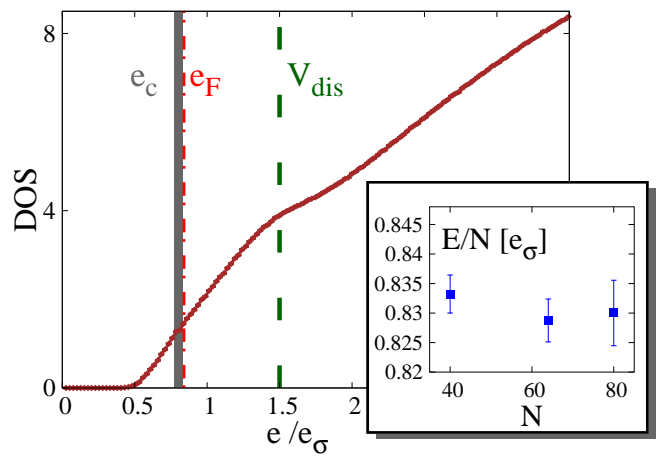}
\caption{(Color online)
Main panel: Single-particle (noninteracting) density of states (in arbitrary units) as a function of the energy $e/e_\sigma$ (brown solid curve). The vertical dashed (green) line indicates the disorder amplitude $V_{\mathrm{dis}}/e_\sigma=1.5$, the vertical (grey) bar the mobility edge $e_{\mathrm{c}}/e_\sigma$, and  the vertical  dot-dashed (red) line the Fermi energy $\epsilon_F$ at the density $n\cong0.185(\pi\sigma)^{-3}$.
Inset: Many-body ground-state energy per particle $E/N$ versus the particle number $N$; the density is $n\cong0.185(\pi\sigma)^{-3}$; the interaction parameter is $k_Fa=0.7$, corresponding to $a\cong0.397\pi\sigma$.
}
\label{fig2}
\end{center}
\end{figure}
%
%

In this Rapid Communication, we analyze the zero-temperature ferromagnetic behavior of the disordered repulsive Fermi gas, determining the critical interaction strength for the Stoner instability in the presence of disorder. We address both the transition from the paramagnetic phase to the partially ferromagnetic phase, and the one from the partially ferromagnetic to the fully ferromagnetic phase (in the case of globally balanced populations). The gas density and the disorder amplitude are tuned so that the Fermi energy of the noninteracting (balanced) gas is close to the mobility edge. This (somewhat arbitrary) choice is motivated by the fact that close to the mobility edge the single-particle orbitals display multifractal properties, a  feature which is expected to enhance the interaction effects~\cite{kravtsov}.
In order to figure out the ferromagnetic behavior, we compute the zero-temperature equation of state as a function of the interaction strength and of the population imbalance, we extract the spin susceptibility, and we determine the energy of a single impurity immersed in a single-component Fermi gas.
Our findings indicate that these quantities are drastically affected by the disorder, displaying a different dependence as a function of the interaction strength compared to the clean gas. More specifically, the magnetic susceptibility is suppressed by the disorder if the repulsion is weak, but it increases with the interaction strength much more rapidly than in the clean gas.  The critical interaction strength where it diverges - which signals the instability towards the partially ferromagnetic phase - is much smaller than in the absence of disorder. The polaron energy is also strongly influenced by disorder, and the critical interaction strength at which it exceeds the chemical potential of the majority component - which signals the transition to the fully ferromagnetic phase - is significantly weaker than in the clean gas. These results indicate that disorder strongly favors the onset of ferromagnetic behavior.\\

\begin{figure}
\begin{center}
\includegraphics[width=1.0\columnwidth]{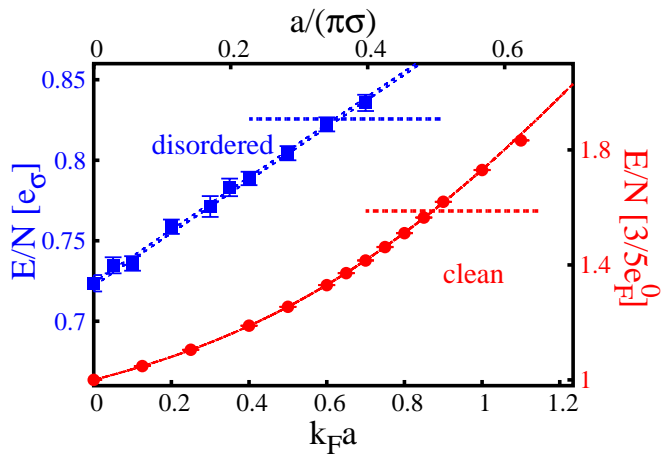}
\caption{(Color online)
Energy per particle $E/N$ at polarization $P=0$ as a function of the scattering length $a/(\pi\sigma)$ for the clean gas (red circles, right vertical axis) and for the disordered gas (blue squares, left vertical axis) with disorder amplitude $V_{\mathrm{dis}}=1.5e_\sigma$. The interaction strength can be expressed also as $k_Fa$, where the Fermi wave-vector $k_F= (3\pi^2n)^{1/3}$ is defined with the average density $n\cong0.185(\pi\sigma)^{-3}$. 
The solid (red) curve is a fourth-oder fit to the DMC data, the dashed (blue) line is a linear fit (see text). The horizontal dashed segments indicate the corresponding energies of the fully imbalanced gases. $3/5e_F^0$ is the energy per particle of a clean noninteracting Fermi gas.
}
\label{fig3}
\end{center}
\end{figure}
 %
%
The disordered Fermi gas we consider is described by the  following Hamiltonian:
\begin{equation}
H = \sum_{\sigma=\uparrow,\downarrow}
       \sum_{i_\sigma=1             }^{N_\sigma      }\left(-\Lambda\nabla^2_{i_\sigma}  + V(\mathbf{r}_{i_\sigma}    )\right)
       +  \sum_{i_\uparrow,i_\downarrow}v(r_{i_\uparrow i_\downarrow}) 
       \;;
\label{hamiltonian}
\end{equation}
here, $m$ is the atomic mass, $\hbar$ is the reduced Planck constant, and we introduced $\Lambda=\hbar^2/2m$. The indices $i_\uparrow$ and $i_\downarrow$ label atoms of the two species, hereafter referred to as spin-up and spin-down particles. The distance between unlike fermions is $r_{i_\uparrow i_\downarrow} = \left|\mathbf{r}_{i_\uparrow}-\mathbf{r}_{i_\downarrow}\right|$. The total number of fermions is $N=N_\uparrow+N_\downarrow$, and the polarization is defined as $P=(N_\uparrow-N_\downarrow)/N$.
The system is enclosed in a cubic box of size $L$ with periodic boundary conditions.
$v(r)$ is a model potential that describes the short-range (inter-species) interactions. In a sufficiently dilute and cold gas, the interaction strength is parametrized just by the $s$-wave scattering length $a$ (this parameter can be tuned experimentally using Feshbach resonances~\cite{chin}), while the other details of the inter-atomic potential as, e.g., the effective range $r_{\mathrm{eff}}$ and and the p-wave scattering length $a_p$, are irrelevant.
We choose the hard-sphere model: $v(r)=+\infty$ if $r<R_0$ and zero
otherwise; in this case, one has $a=R_0$, $r_{\mathrm{eff}}=2a/3$, and $a_p=a$.
The possible nonuniversal effects due to the details beyond $a$ have been thoroughly analyzed in Refs.~\cite{pilati2010,trivedi,polls} and in Ref.~\cite{pilati2014} - for homogeneous gases and for non-homogenous gases exposed to periodic potentials, respectively - using different models for the interatomic interactions, including the zero-range pseudopotential.
Both in the homogeneous and in the non-homogeneous case, it was found that the equation of state (and, hence, the critical interaction strength for the Stoner instability) is affected by about $10\%$ in the strong interaction regime $k_Fa \gtrsim 1$, where $k_F = (3\pi^2 n)^{1/3}$ is the Fermi wave-vector defined with the average density $n=N/L^3$, and that these nonuniversal effects rapidly vanish for weaker interactions. In this Rapid Communication, we only consider the moderate interaction regime $k_Fa \lesssim 1$, where the nonuniversal effects do not play a significant role.\\
$V(\bf{r})$ is an external random field that describes the effect due to a blue-detuned isotropic optical speckle pattern. In cold-atoms experiments, speckle patterns are realized by shining lasers through diffusive plates, and then focusing the diffused light onto the atomic cloud~\cite{aspect2009anderson,sanchez2010disordered}.
In the case of a blue-detuned optical field, the atoms experience a repulsive potential with the exponential local-intensity distribution: $P_{\textrm{bd}}(V) = \exp\left(-V/V_{\mathrm{dis}}\right)/V_{\mathrm{dis}}$, if the local intensity is $V\geqslant0$, and $P_{\textrm{bd}}(V)=0$ otherwise~\cite{goodman1975statistical}. 
The  parameter $V_{\mathrm{dis}}\geqslant 0$ fixes both the spatial average of the random field $V_{\mathrm{dis}}=\left<V(\bf{r})\right>$ and its standard deviation, so that: $V_{\mathrm{dis}}^2=\left<V({\bf r})^2\right>-\left<V({\bf r})\right>^2$; therefore, $V_{\mathrm{dis}}$ is the parameter that characterizes the global disorder amplitude.
The two-point spatial correlations of the speckle field depend on the profile of the illumination on the diffusive plate and on the details of the optical setup. We consider the idealized case where the spatial correlations are isotropic, being described by the following correlation function~\cite{delande2014mobility}: $\Gamma(r=|{\bf r}|) = \left< V({\bf r}'+{\bf r}) V({\bf r}')\right>/V_{\mathrm{dis}}^2-1= \left[\sin(r/\sigma)/(r/\sigma)\right]^2$ (here we assume averaging over the position of the first point ${\bf r'}$). 
The parameter $\sigma$ determines the length scale of the spatial correlations. The correlation function $\Gamma(r)$ rapidly decreases with the distance $r$; it vanishes at the distance $r=\pi\sigma$, which corresponds to the typical speckle size and, thus, to the disorder correlation length; for larger distances $\Gamma(r)$ displays small oscillations. To favor comparison with previous literature, we will express length scales in unit of the correlation length $\pi\sigma$, and the energy scales in unit of the correlation energy $e_{\sigma} = \hbar^2/(m\sigma^2)$.
In our simulations, the isotropic speckle pattern is generated following the numerical recipe described in Ref.~\cite{fratini1}; it satisfies the periodic boundary conditions. See also Refs.~\cite{huntley1989speckle,modugno2006collective,delande2014mobility}.\\
 %
\begin{figure}
\begin{center}
\includegraphics[width=1.0\columnwidth]{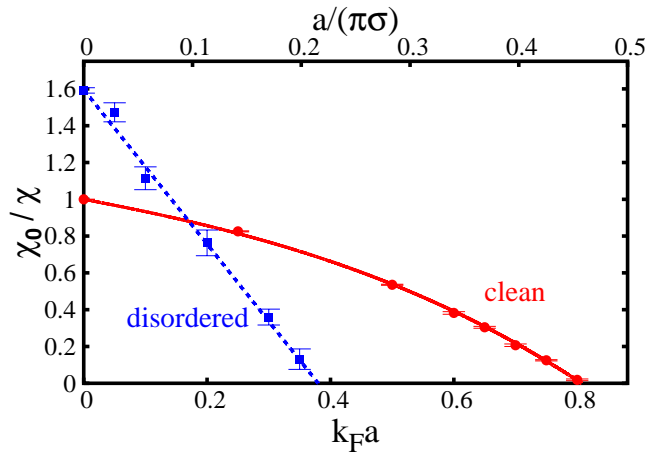}
\caption{(Color online) Inverse magnetic susceptibility $\chi_0/\chi$ for the clean gas (red circles) and for the disordered gas (blue squares). $\chi_0=3n/(2e_F)$ is the result corresponding to the clean noninteracting Fermi gas. The solid (red) curve is a cubic fit to the DMC data, the dashed (blue) line is a linear fit (see text).
}
\label{fig4}
\end{center}
\end{figure}

To determine the ground-state properties of the Hamiltonian~(\ref{hamiltonian}) we employ quantum Monte Carlo simulations based on the fixed-node diffusion Monte Carlo (DMC) algorithm~\cite{reynolds1982fixed}. The DMC algorithm is designed to sample the lowest-energy wave function by stochastically evolving the Schr\"odinger equation in imaginary time. The fixed-node constraint - which consists in imposing that the nodal surface of the many-body wave-function is the same as that of a trial wave function $\psi_T$ - is introduced in order to circumvent the sign problem, which would otherwise hinder fermionic Monte Carlo simulations.
If the nodal surface of $\psi_T$ is exact, this variational method provides unbiased estimates of the ground-state energy. In general, the predicted energies are rigorous upper bounds, which have been found to be very close to the exact ground state energy if the nodes of $\psi_T$ are good approximations of the ground-state nodal surface (see, {\it e.g.}, \cite{foulkes}). 
We adopt trial wave functions of the Jastrow-Slater type:
\begin{equation}
\psi_T({\bf R})= D_\uparrow(N_\uparrow) D_\downarrow(N_\downarrow) \prod_{i_\uparrow,i_\downarrow}f(r_{i_\uparrow i_\downarrow}) \;,
\label{psiT}
\end{equation}
where ${\bf R}=({\bf r}_1,..., {\bf r}_N)$ is the spatial configuration vector and $D_{\uparrow(\downarrow)}$ denotes the Slater determinant of single-particle orbitals of the particles with up (down) spin. The Jastrow-Slater trial wave-function has been found to describe very accurately the normal (non superfluid) phases of strongly interacting fermions; for a review on this issue, see Refs.~\cite{foulkes,wagner2016discovering}. In the case of weakly-repulsive atomic Fermi gases in deep optical lattices, which can be described with the single-band Hubbard model derived within a tight-binding scheme, the energies obtained from continuous-space DMC simulations based on the Jastrow-Slater trial wave-function~\cite{pilati2010} have been found to precisely agree with the accurate Hubbard model simulations performed using the constrained path Monte Carlo method~\cite{zhang}.\\
%
 In this study, the Jastrow correlation term $f(r)$ is equal to the solution of the s-wave radial Schr\"odinger equation describing the scattering of two hard-sphere particles in free space, as in Ref.s~\cite{pilati2010,pilati2014}. The scattering energy is fixed by the boundary condition $f^\prime(r=L/2)=0$ on its derivative.
Since $f(r)>0$, the nodal surface is fixed by the antisymmetry of the Slater determinants only. This, in turn, is fixed by the choice for the single-particle orbitals.
In this study, we employ the $N_\uparrow$ ($N_\downarrow$) lowest-energy single-particle eigenstates $\phi_{j}(\bf{r})$ (with $j=0,\dots,N_{\uparrow(\downarrow)}-1$) of the disordered potential $V({\bf r})$, for the spin-up (spin-down) particles. These eigenstates satisfy the equation $\left[-\Lambda \nabla^2  +V({\bf r})\right]\phi_{j}(\mathbf{r}) = e_j \phi_{j}(\mathbf{r})$, with the eigenvalues $e_j$. We determine them via exact numerical diagonalization of the finite matrix obtained after introducing a discretization in the continuous-space and approximating the Laplacian using high-order finite-difference formulas. We carefully analyze how the discretization error affects both the single-particle eigenvalues and the many-body ground-state energy obtained from the DMC simulations, ensuring that the discretization error is negligible compared to the statistical uncertainty of the Monte Carlo predictions. 
Notice that the energy on a noninteracting disordered gas with $N_\uparrow$ spin-up particles and $N_\downarrow$ spin-down particles can be computed as $E(N_\uparrow,N_\downarrow)=\sum_{j=0}^{N_\uparrow-1}e_j + \sum_{j=0}^{N_\downarrow-1}e_j$.\\
%
%
In Fig.~\ref{fig1}, three representative single-particle orbitals are visualized by showing the corresponding three-dimensional probability density distribution $\left|\phi_j({\bf r})\right|^2$. The first is the ground-state ($j=0$), which is localized in a restricted region of space due to the Anderson localization phenomenon. The second is a critical orbital corresponding to an energy near the mobility edge $e_c$; it has an intermediate character between extended and localized. The third orbital corresponds to an eigenvalue well above the mobility edge (and the Fermi energy). It is worth emphasizing that the one-particle density distribution corresponding to the many-fermion wave-function will likely be much more homogeneous than the one corresponding to individual single-particle orbitals.\\
%

\begin{figure}
\begin{center}
\includegraphics[width=1.0\columnwidth]{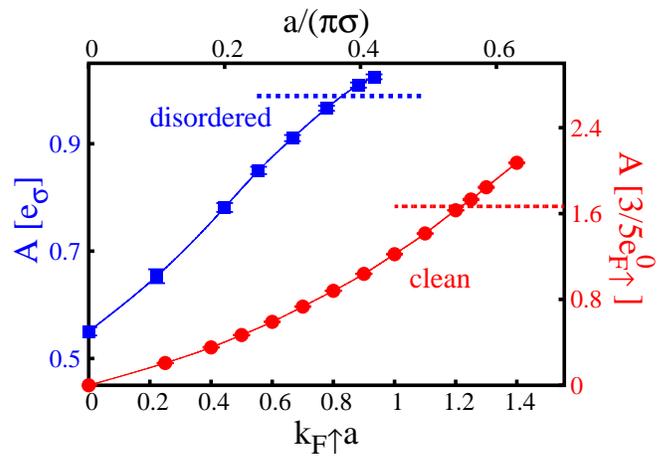}
\caption{(Color online)  Chemical potential at zero concentration of the repulsive polaron in the clean gas (red circles, right axis) and in the disordered gas (blue squares, left axis). $e_{F\uparrow}^0=(\hbar k_{F\uparrow})^2/2m$ is the Fermi energy of the clean fully-imbalanced noninteracting Fermi gas. The interaction parameter $a/(\pi\sigma)$ can be cast in the form  $k_{F\uparrow}a$ using the Fermi wave-vector $k_{F\uparrow}=\left(6\pi^2n_\uparrow\right)^{1/3}$ defined with the average spin-up density $n_\uparrow=N_\uparrow/L^3\cong0.185(\pi\sigma)^{-3}$. The horizontal dashed segments indicate the chemical potential of the majority component. The solid curves through DMC data are guides to the eye.}
\label{fig5}
\end{center}
\end{figure}

In this Rapid Communication, we first determine the zero-temperature equation of state of a population balanced Fermi gas with $N_\uparrow=N_\downarrow$ ($P=0$).
 We consider a blue-detuned isotropic speckle field with average intensity $V_{\mathrm{dis}}=1.5e_{\sigma}$. The density of states of the noninteracting problem at this disorder strength is shown in Fig.~\ref{fig2} (main panel). The mobility edge $e_{\mathrm{c}}$, namely the energy  threshold that separates the localized single-particle orbitals with energies $e<e_{\mathrm{c}}$ from the extended orbitals with energies $e>e_{\mathrm{c}}$, is $e_{\mathrm{c}}=0.80(3)e_{\sigma}$. This value is obtained from the statistical analysis of the spacings between energy levels~\cite{haake2010quantum}, exploiting the universal value of the critical adjacent level-spacings ratio, following the procedure of Refs.~\cite{fratini1,fratini2}. This result agrees within statistical errors with the prediction $e_{\mathrm{c}}=0.787(9)e_{\sigma}$ obtained in Ref.~\cite{delande2014mobility} using the transfer matrix method. We consider a gas with fixed average density $n\cong0.185(\pi\sigma)^{-3}$, for which the (noninteracting) Fermi energy is found to be $e_F \simeq 0.84e_{\sigma}$, just above the mobility edge.
 %
 %
 In Fig.~\ref{fig3} we plot the energy per particle $E/N$ computed using the DMC algorithm described above, for increasing values of the s-wave scattering length $a$. We stress that in this study the adimensional density parameter $n(\pi\sigma)^3$ is fixed, while the ratio $a/(\pi\sigma)$ increases. The interaction parameter can also be cast in the form $k_F a$, familiar from the theory of clean Fermi gases, defining the Fermi wave-vector using the average density $n$. These results have been obtained by averaging over 15 to 20 realizations of the speckle field, using systems sizes from $N=40$ to $N=80$. The analysis of the finite-size effects (shown in the inset of Fig.~\ref{fig2}) suggests that the systems we simulate are sufficiently large to predict the ground-state energies corresponding to the thermodynamic limit, within the statistical uncertainties.
 The ground-state energy $E/N$ displays a different dependence on the interaction strength with respect to the clean gas (also shown in Fig.~\ref{fig3}, data from Ref.~\cite{pilati2010}). In the latter case, the equation of state is well described by the polynomial $E/N=e_u\sum_{i=0,\dots,4} c_i (k_F a)^{i}$,  where $c_0=1$, $c_1\cong0.3536$, and $c_2\cong0.1855$ are provided by the second-order perturbation theory~\cite{PhysRev.105.767,PhysRev.105.1119}, while $c_3=0.307(7)$ and $c_4=-0.115(8)$ are fitting parameters~\cite{dft}, and the energy unit is $e_u=3/5e_F^0$, namely the energy per particle of the clean noninteracting Fermi gas (being $e_F^0=(\hbar k_F)^2/(2m)$ the corresponding Fermi energy).
 In the disordered gas, a simple linear form (with $c_0=0.723(2)$, $c_1 = 0.164(5)$, $e_u = e_\sigma$, and $c_i=0$ for $i=2,3,4$) appears to accurately fit the data.
 It is surprising to observe this simple behavior emerging in a such a complex quantum systems.
We observe that in the disordered gas the energy of the balanced ($P=0$) gas exceeds the energy of the fully imbalanced gas with $P=1$ (which is indicated with an horizontal segment in Fig.~\ref{fig3}) at weaker interaction strength than in the absence of disorder. Exceeding the fully-imbalanced gas energy is a sufficient - though, not necessary- condition for the occurrence of ferromagnetic behavior~\cite{trivedi}. This finding already suggests that disorder favors the onset of ferromagnetism in repulsive Fermi gases. A more precise characterization of the ferromagnetic properties can be obtained by analyzing the spin susceptibility and the polaron energy, as described below.\\

The transition from the paramagnetic phase to the partially ferromagnetic phase can be identified from the divergence of the spin susceptibility $\chi=n\left(\frac{\partial^2 (E/N)}{\partial P^2}\right)^{-1}$. This criterion is associated to a second-order phase transition. We determine $\chi$ by performing DMC simulations with imbalanced populations with $N_\uparrow>N_\downarrow$ ($0\leqslant P\leqslant 1$), keeping the total particle number $N$ fixed, for individual disorder realizations. For sufficiently small values of the population imbalance $P$, the energy per particle is found to vary with $P$ according to the quadratic function: $E(P)/N = E(P=0)/N + nP^2/(2\chi)$, where $E(P=0)$ and $\chi$ are fitting parameters.
 In Fig.~\ref{fig4} we show the inverse susceptibility $\chi^{-1}$ as a function of the interaction strength, obtained after averaging over 5 to 10 disorder realizations. We notice that in the weakly interacting limit, $\chi^{-1}$ is larger than in the clean gas (data from Ref.~\cite{dft}), meaning that the disorder alone (i.e., in the absence of interactions) suppresses the spin fluctuations. However, $\chi^{-1}$ quickly drops to zero as the interaction parameter increases, indicating a strong interplay between disorder and interactions. Already at the interaction strength $k_Fa\simeq0.2$, the inverse susceptibility is smaller in the disordered gas than in the clean gas. In the disordered case, the critical point where $\chi^{-1}$ vanishes, which signals the transition to the partially ferromagnetic phase, is $k_F a\simeq0.38$, considerably smaller than the corresponding value for the clean gas  $k_F a\simeq0.80$~\cite{dft}. We point out that, while these findings support the scenario of a second-order phase transition, our numerics cannot rule out an extremely weakly first-order transition~\cite{PhysRevLett.82.4707}, and they are also consistent with the infinite-order transition proposed in Ref.~\cite{carleo}. Furthermore, in the case of clean systems, more exotic magnetic phases with spin-textured magnetization have been predicted to occur in the close vicinity of the ferromagnetic transition~\cite{conduit}; we do not consider these spin-textured phases.
 We also notice that, while in the clean gas the inverse susceptibility is well described by the cubic fitting function $\chi_0/\chi = \sum_{i=0,\dots,3} d_i (k_F a)^{i}$, where $d_0=1$, $d_1\cong-0.637$, and $d_2\cong-0.291$ are provided by perturbation theory~\cite{recati2011spin}, and $d_3=-0.56(1)$ is a fitting parameter, in the disordered case the simple linear fitting function (with $d_0=1.60(1)$, $d_1\cong-4.2(1)$, and $d_i=0$ for $i=2,3$) precisely reproduces the trend of the data.\\
 The transition from the partially ferromagnetic phase to the fully ferromagnetic phase can be located by determining the chemical potential at zero-concentration of the repulsive polaron, defined as $A=E(N_\uparrow,1)-E(N_\uparrow,0)$, where $E(N_\uparrow,1)$ is the energy of a gas with $N_\uparrow$ spin-up particles plus a spin-down impurity, and $E(N_\uparrow,0)$ is the energy of the $N_\uparrow$ spin-up particles alone. For sufficiently strong repulsion, $A$ exceeds the chemical potential of the majority component (which we compute as $e_{F\uparrow}=E(N_\uparrow+1,0)-E(N_\uparrow,0)$). At this point, the fully ferromagnetic phase becomes energetically favorable, meaning that the configuration with fully separated domains, each hosting particles of one species only, is stable~\cite{pilati2010,cui}. The data shown in Fig.~\ref{fig5} (obtained by averaging 10 to 20 disorder realizations) indicate that in the presence of disorder this transition takes place at $k_{F\uparrow} a\simeq 0.82$ (where $k_{F\uparrow}=(6\pi^2n_\uparrow)^{1/3}$ is the majority-component Fermi wave-vector, defined with the corresponding average density $n_\uparrow=N_\uparrow/L^3$), which is significantly smaller than in the clean gas: $k_{F\uparrow} a \simeq 1.22$. 
It is worth noticing that, while in the absence of disorder the partially ferromagnetic phase is stable in the window $0.80 < k_Fa < 0.97$~\cite{pilati2010,dft}, in the presence of disorder this window is shifted to weaker interactions, and it is enlarged: $0.38 < k_Fa < 0.65$ (we used the conversion $k_F=k_{F\uparrow}/2^{1/3}$).\\

 In conclusion, we investigated the zero-temperature properties of disordered interacting Fermi gases, including the equation of state, the magnetic susceptibility, and the polaron energy. We employed quantum Monte Carlo simulations of realistic models that describe the disorder due to optical speckle patterns. We analyzed the interplay between the Stoner ferromagnetic instability and the Anderson localization transition.
 We observed that disorder strongly favors the onset of ferromagnetic behavior, shifting the ferromagnetic transitions to significantly weaker interaction than in the clean gas. These results suggest a new paradigm to explain the emergence of ferromagnetic behavior, extending beyond the case of itinerant (delocalized) fermions, and indicate an alternative route to observe quantum magnetism is cold-atoms experiments, circumventing the molecule-formation problem that plagues the regime of strong interactions. The data we provide constitute also as a valuable benchmark to develop new theories for the Fermi glass regime~\cite{freedman,fleishman}.\\

We thank Alessio Recati and Giacomo Roati for useful discussions and for illustrating to us the results of Ref.~\cite{roati}.


%

\end{document}